\documentclass[%
 aip,
rsi,%
 amsmath,amssymb,floatfix,
 a4paper,
 reprint,%
]{revtex4-1}

\usepackage{graphicx}
\usepackage{dcolumn}
\usepackage{bm}
\usepackage{float}
\usepackage[english]{babel}
\usepackage{xcolor}

\usepackage{color,soul}

\usepackage[slantedGreek]{newtxmath}
\newcommand{\uscript}[1]{$_{\text{#1}}$}

\begin{document}

\preprint{ArXiv}

\title[Stable and compact RF-to-optical link using lithium niobate on insulator waveguides]{\large{Stable and compact RF-to-optical link using lithium niobate on insulator waveguides} \vspace{0.4cm}}

\author{Ewelina Obrzud}
\author{Séverine Denis}
\author{Hamed Sattari}
\author{Gregory Choong}
\author{Stefan Kundermann}
\author{\\ Olivier Dubochet}
\author{Michel Despont}
\author{Steve Lecomte}
\author{Amir Ghadimi}
\author{Victor Brasch \vspace{0.2cm}}
\email{victor.brasch@csem.ch}

\begin{abstract}
\textit{Centre Suisse d'Electronique et de Microtechnique (CSEM), 2000 Neuchâtel, Switzerland} \vspace{0.3cm}
 \vspace{0.1cm}
 Optical frequency combs have become a very powerful tool in metrology and beyond thanks to their ability to link radio frequencies with optical frequencies via a process known as self-referencing. Typical self-referencing is accomplished in two steps: the generation of an octave-spanning supercontinuum spectrum and the frequency-doubling of one part of that spectrum. Traditionally, these two steps have been performed by two separate optical components. With the advent of photonic integrated circuits, the combination of these two steps has become possible in a single small and monolithic chip. One photonic integrated circuit platform very well suited for on-chip self-referencing is lithium niobate on insulator -- a platform characterised by high second and third order nonlinearities. Here we show that combining a lithium niobate on insulator waveguide with a silicon photodiode results in a very compact and direct low-noise path towards self-referencing of mode-locked lasers. Using digital servo electronics the resulting frequency comb is fully stabilized. Its high degree of stability is verified with an independent out-of-loop measurement and is quantified to be 6.8\,mHz. Furthermore, we show that the spectrum generated inside the lithium niobate waveguide remains stable over many hours.
 \end{abstract}

\maketitle
\thispagestyle{empty}

\section{Introduction}
Lithium niobate on insulator (LNOI) has gained a lot of momentum in recent years as a platform for photonic integrated circuits (PICs)\cite{boes_2018, Zhu2021}. Impressive demonstrations include for example efficient electro-optic modulators \cite{wang_integrated_2018, hu_folded_2021, kharel_breaking_2021}, an on-chip spectrometer \cite{Pohl2019}, lasers and amplifiers \cite{Yin2021, Wang2021, Chen2021}, optical parametric oscillators \cite{Juanjuan2021, mckenna_ultra-low-power_2021}, efficient second harmonic generation \cite{chen_ultra-efficient_2019, lu_toward_2020} , acousto-optic modulation \cite{sarabalis_acousto-optic_2021}, THz generation \cite{Yang2021, Carnio2017}, Kerr frequency combs \cite{Wang2019, He2019_microcomb, Gong2020}, supercontinuum generation \cite{lu_octave-spanning_2019, jankowski_ultrabroadband_2020, Escale2020}  and self-referencing of frequency combs \cite{yu_coherent_2019, okawachi_chip-based_2020}. These results show the versatility of the LNOI platform and its great potential for future applications in fields such as sensing, distance measurement, communication, metrology and many more \cite{Zhu2021}. This interest and success stems from lithium niobate's excellent optical properties: a wide  optical transmission window ranging from 350\,nm to 5\,$\muup$m and high second and third order nonlinearities as well as electro-optic and piezo-electric properties. What adds to the platform's appeal is the fact that LNOI is already commercially available as wafers of high quality monocrystalline thin films on silicon dioxide. 

\begin{figure}[!t]
\centering
\includegraphics[width=1\linewidth]{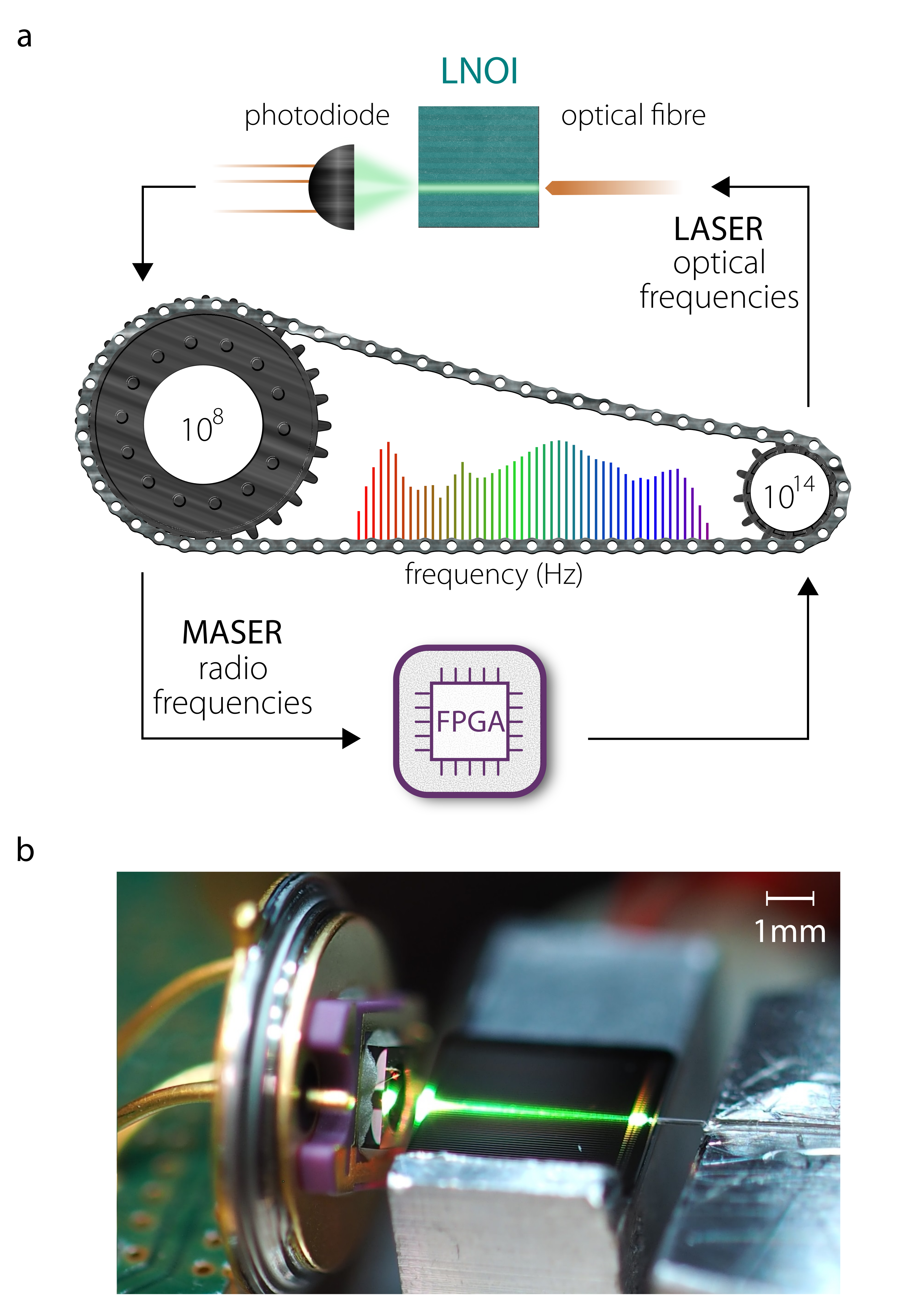}
\caption{\textbf{Optical-to-RF link.} \textbf{a} A lensed optical fibre couples the optical femtosecond pulses to a lithium niobate on insulator (LNOI) waveguide. The nonlinear optical processes inside the waveguide bridge from the optical to the RF domain. The carrier-envelope offset frequency f\uscript{ceo} of the optical frequency comb becomes detectable  in the RF domain with a photodiode directly after the waveguide. The digital feedback onto the laser via a field programmable gate array (FPGA) completes the link between RF and optical domain. \textbf{b} A photograph of the optical-to-RF link in operation. From the right: optical input fibre, LNOI PIC with waveguide (glowing green) and a Si PIN photodiode.}
\label{fig:principle}
\end{figure}

One application where LNOI PICs have been demonstrated a large potential for a reduction in size, cost and complexity is the self-referencing of frequency combs \cite{yu_coherent_2019, okawachi_chip-based_2020}. Traditionally, laser stabilisation is achieved with a combination of fiber optics and bulk optics, where separate optical components are used for supercontinuum and harmonic generation. On the contrary, self-referencing with LNOI can provide a considerable simplification due to the coexisting high second and third order nonlinearities in lithium niobate. By injecting femtosecond pulses into a LNOI waveguide, a spectral overlap between the generated supercontinuum and the simultaneously generated second harmonic can be achieved leading to a detection of f\uscript{ceo}, the carrier-envelope offset frequency\cite{udem_optical_2002}. This direct on-chip f-2f interferometry can provide the crucial element of a compact and efficient RF-to-optical link, as shown in Figure \ref{fig:principle}a. 

Additionally, two advantages arise when using PICs for applications relying on nonlinear frequency conversion: first, compared to optical fibers, the required pulse energies for achieving coherent octave-spanning spectra are relatively low, in the range of several tens of picojoules, due to the tight light confinement in the PICs' waveguides and the often high intrinsic optical nonlinearities of the waveguide materials. Second, the effectiveness of the nonlinear processes is increased by waveguide dispersion engineering, allowing a careful control over the dispersion curves and obtaining a desired result suitable for a given application. This has been demonstrated over the last years for several different PIC platforms\cite{halir_ultrabroadband_2012, liu_beyond_2019, obrzud_visible_2019, Zhang2019, singh_supercontinuum_2019, lu_octave-spanning_2019, jankowski_ultrabroadband_2020, kuyken_octave-spanning_2020, woods_supercontinuum_2020, Escale2020}. All these properties, in conjunction with demonstrated techniques for achieving low input coupling losses \cite{He2019_low_loss, Ying2021, Hu2021}, add to the great appeal of the LNOI PIC platform.

In this article, we address the suitability of LNOI PICs for long-term operation and the metrological stability of the laser lock. We show that a millimeter-sized LNOI PIC combined with a standard silicon PIN photodiode and fully digital servo electronics allows for an extremely simple yet very stable operation of a self-referenced and fully locked frequency comb. To prove the actual metrological stability of the system, we build an out-of-loop setup as an independent fibre-based f-2f interferometer, which demonstrates a millihertz-level stability. In addition, we investigate the stability of the optical supercontinuum spectra generated inside the LNOI waveguides. The shape of the optical spectrum does not show any significant changes over many hours of continuous operation and power variations in the spectrum stay well below 25$\%$.   

\begin{figure*}[!t]
\centering
\includegraphics[width=1\textwidth]{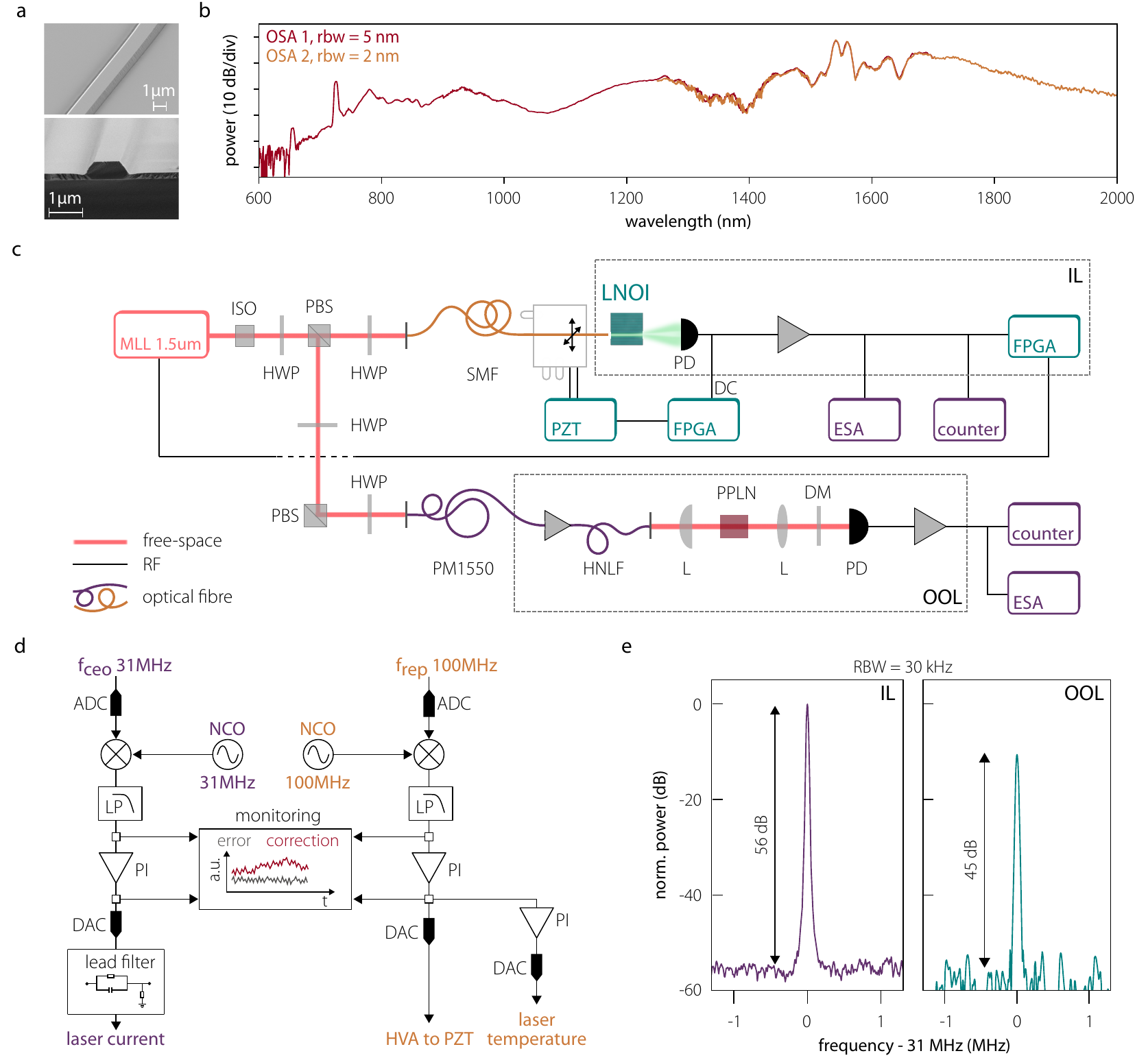}
\caption[width=1\textwidth]{\textbf{Optical spectrum, experimental setup,  and f\uscript{ceo} in-loop and out-of-loop beatnotes.} \textbf{a} Representative scanning electron microscopy images of lithium niobate on insulator waveguides (top) and their cross section (bottom) fabricated at CSEM. \textbf{b} Optical spectrum after a LNOI waveguide with a nominal width of 1300\,nm illuminated with $\sim$140\,pJ optical pulses centered around 1555\,nm. \textbf{c} Experimental setup for compact laser stabilisation using a LNOI waveguide and the metrological characterisation of the laser lock. LNOI: lithium niobate on insulator PIC, MLL: mode-locked laser, ISO: optical isolator, PBS: polarising beam splitter, HWP: half-wave plate, SMF: single-mode fibre, PD: photodiode, FPGA: field programmable gate array, PZT: piezo controller, ESA: electrical spectrum analyser, PM1550: polarisation-maintaining single mode fibre, HNLF: highly nonlinear fibre, L: lens, DM: dichroic mirror, PPLN:  periodically-poled lithium niobate crystal, IL: in-loop, OOL: out-of-loop. \textbf{d} Laser lock scheme. ADC: analog-to-digital converter, NCO: numerically-controlled oscillator, LP: low-pass filter, PI: proportional-integral gain, DAC: digital-to-analog converter, HVA: high voltage amplifier. \textbf{e} Measured in-loop (left) and out-of-loop (right)  f\uscript{ceo} beatnotes.\vspace{0.3cm}}
\label{fig:setup}
\end{figure*} 

\section{LNOI PIC Fabrication}

The LNOI PICs with the nonlinear waveguides used in this study are fabricated at CSEM's micro-fabrication facility. The waveguide fabrication technology is based on commercially available LNOI 6-inch wafers (from NanoLN). The thin film layer stack consists of a 600\,nm thick mono-crystalline x-cut lithium niobate layer on top of a 4.7\,$\muup$m buried thermal oxide layer. The waveguides are patterned in hydrogen silsesquioxane (HSQ) using electron beam lithography and are etched 400\,nm into the LNOI layer using an optimized Ar+ ion milling recipe \cite{zhang2017monolithic} that results in smooth sidewalls with an angle of ~ 35° with respect to normal (Fig.\ref{fig:setup}a). For this experiment, the waveguides have an air top cladding to achieve the dispersion regime  required for octave spanning supercontinuum generation. Finally the individual PICs have been released from the wafer using a reactive ion etching (RIE) deep silicon etching process that results in smooth PIC sidewalls that are critical for efficient edge coupling. 

\section{Experimental Setup}

For the experimental implementation of the laser self-referencing we use a LNOI waveguide whose nominal width is 1300\,nm. Propagating $\sim$ 140\,pJ optical pulses centred around 1555\,nm through the waveguide results in a more than octave-spanning optical spectrum, as shown in Fig.\ref{fig:setup}b. The emission from the waveguide is characterised by strong second harmonic (SH) generation in the wavelength range from around 600\,nm to 900\,nm, partially overlapping with the supercontinuum (SC). A substantial amount of third harmonic is also generated along the waveguide, as indicated by the green glow of the waveguide during operation (Fig.\ref{fig:principle}b). To demonstrate the potential of such a setup as a simple and compact self-referencing stage, the output light from the LNOI waveguide is directly detected with a silicon photodiode to measure the offset frequency. 

The setup, presented in Fig.\ref{fig:setup}c, consists of a solid-state mode-locked laser operating at 1555\,nm delivering 150\,fs optical pulses with an average power of up to 140\,mW at 100\,MHz repetition rate. The laser light is divided into two parts. The first part is sent to the lithium niobate waveguide to provide the on-chip self-referencing (in-loop, IL). In order to achieve this, approximately 60\,mW of average power is coupled into a standard single-mode lensed fibre (SMF) with its end mounted on a 3-axis stage. The pulse propagation along the SMF results in a slight temporal pulse compression, down to approximately 125\,fs. The lensed optical fibre is aligned for optimal light injection into the waveguide on the LNOI PIC where pulses undergo nonlinear frequency conversion. The overlap between the SC and SH results in a f\uscript{ceo} beatnote that is detected with a silicon PIN photodiode. Without any optical filtering, the photodiode is placed directly at the waveguide output. The RF signal from the photodiode is split into a DC and an RF part, the former being used to actively stabilise the input optical fibre position. The RF signal, corresponding to f\uscript{ceo} at 31\,MHz, is amplified, split for monitoring and finally recorded by a field-programmable gate array (FPGA) to be used for self-referenced locking of the frequency comb. The f\uscript{ceo} beatnote is monitored by an electrical spectrum analyser (ESA) and the frequency stability is measured with a frequency counter.
In order to assure a stable operation regardless the environmental factors such as the air turbulence and temperature changes, the full optical setup is placed in a protective box and, additionally, the LNOI chip sits on an aluminium mount which is temperature stabilized.

\begin{figure*}[!t]
\centering
\includegraphics[width=1\textwidth]{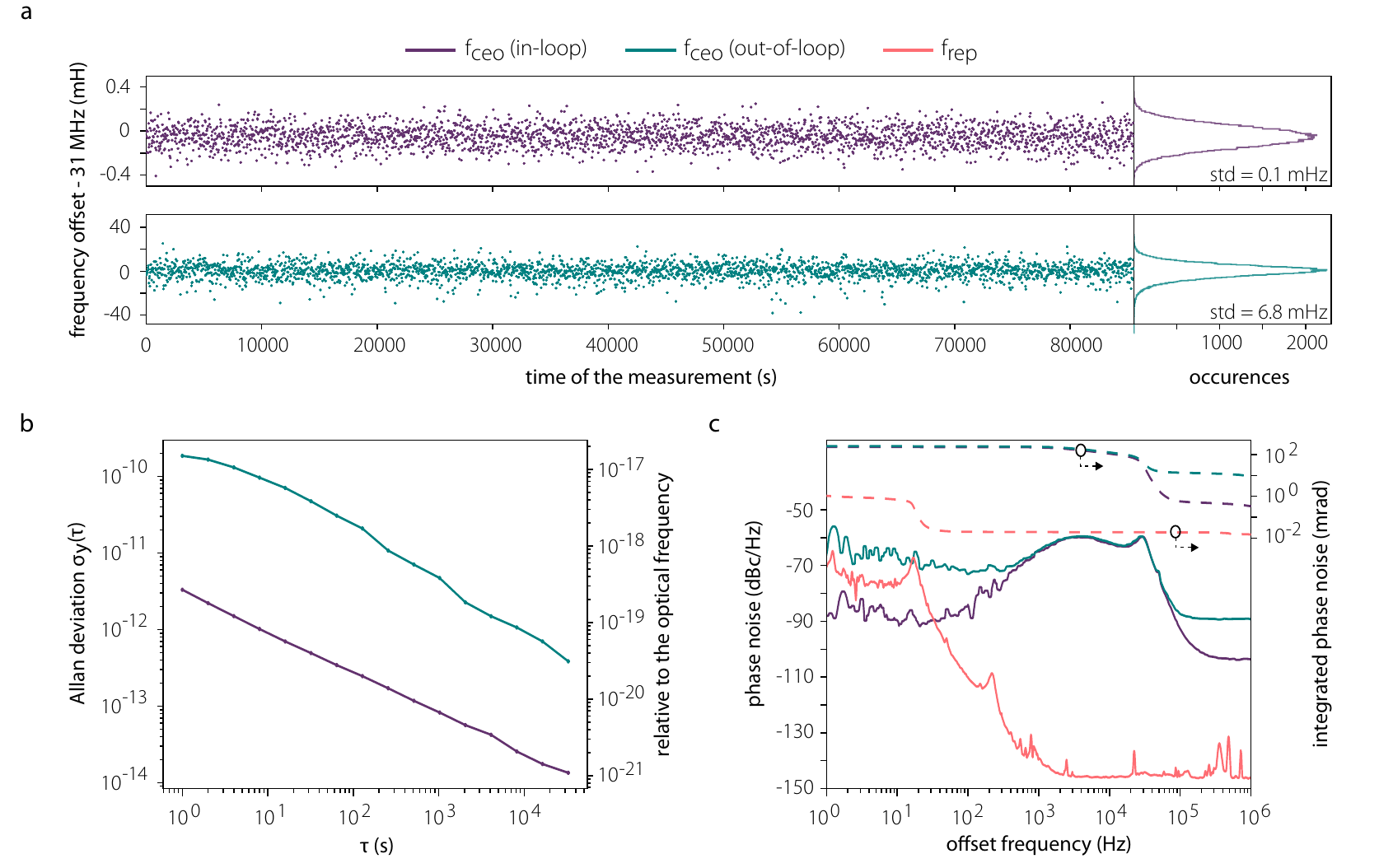}
\caption[width=1\textwidth]{\textbf{Laser lock stability and phase noise.} \textbf{a} f\uscript{ceo} measurement over time (left) and histograms (right) for both the in-loop (top, violet) and out-of-loop (bottom, teal) f\uscript{ceo}. The measurement gate time is 1\,s. The f\uscript{ceo} is offset by 31\,MHz, which was the reference frequency. For clarity, the left panels show every 30th data point only. \textbf{b} Overlapping Allan deviation of the frequency series shown in \textbf{a}. The stability relative to the optical frequency (384\,THz) is also shown. The frequency stability measurement of the in-loop f\uscript{ceo} was limited by the frequency counter. \textbf{c} Representative f\uscript{ceo} and f\uscript{rep} phase noise measurement (solid) and the integrated phase noise (dashed) for a fully stabilised laser.\vspace{0.25cm}}
\label{fig:FreqStab}
\end{figure*} 

The second part of the laser light (several mW, amplified up to 70\,mW in a fiber amplifier) is used for the out-of-loop (OOL) measurement of the f\uscript{ceo} stability by using a standard f-2f interferometer employing a highly nonlinear fibre (HNLF) to generate an octave-spanning supercontinuum and a bulk periodically-poled lithium niobate crystal (PPLN) for second harmonic generation. A dichroic mirror (DM) transmitting the part of the light below 1200\,nm is placed in front of the OOL photodiode in order to avoid saturation. Finally, the resulting OOL f\uscript{ceo} signal is amplified, filtered and monitored on a second ESA and a second frequency counter.  

For a long-term operation it is essential that the optical power coupled to the waveguide stays constant. In our work the input fibre is actively stabilised in two axes by locking the DC signal of the IL PD to its maximum. This is achieved by slowly modulating the two axes describing the plane parallel to the chip facet with the use of piezo actuators and implementing a digital lock-in scheme with a second FPGA.

The scheme of the laser lock is shown in Fig. \ref{fig:setup}d. The stabilization of both f\uscript{rep}, which is detected with a photodiode inside the laser and is for simplicity not shown in Fig.\ref{fig:setup}c, and f\uscript{ceo} is performed on the same FPGA board (RedPitaya SignalLab). Beside the FPGA as part of a system on a chip, this board is equipped with two fast 12\,bit analog-to-digital converters (ADC) and two fast 14\,bit digital-to-analog converters (DAC) with a sampling rate of 250\,MS/s each. The digital part of the stabilization design is implemented using the oscimpDigital ecosystem\cite{oscimpDigital}. In contrast to the FPGA code used in previous demonstrations of frequency combs stabilized with the RedPitaya hardware platform\cite{tourigny-plante_open_2018, shaw_versatile_2019}, oscimpDigital provides a flexible open source tool box suitable also for many other RF applications. The stabilization schemes of f\uscript{rep} and f\uscript{ceo} are quite similar: f\uscript{rep} and f\uscript{ceo} error signals are extracted separately via a digital phase detection between the input signals and two numerical reference oscillators generated inside the FPGA. In both cases a proportional-integral (PI) controller is used to generate the correction signals from the phase error signals. In the case of f\uscript{ceo}, the correction signal is directly sent to the first DAC for a feedback on the pump current of the laser. Meanwhile the correction signal for f\uscript{rep} is duplicated in the FPGA. The first copy is routed to the second DAC and then send to a high voltage amplifier for a “fast” feedback on a piezo-mounted mirror inside the laser cavity. The second copy is averaged and decimated in time to constitute the error signal to compensate the slow drift of f\uscript{rep}. It is then routed to a separate PI controller and the resulting correction signal is converted using a slow auxiliary DAC with a sampling rate of 100\,kS/s. Finally, the signal is sent to the controller of the laser cavity temperature for a slow compensation of the f\uscript{rep} long-term drift. Additionally, an analog lead filter is used to increase the effective electronic feedback bandwidth for f\uscript{ceo}.

Once set up and with the initial adjustments done, the FPGA-based logic for the full stabilization runs entirely on the FPGA board and is independent of any remote computer or network connections. The adjustment of the PI parameters is performed through a web interface and the error and correction signals are streamed through the local network to be monitored and recorded on a remote computer using a GNU Radio Companion \cite{gnuRadio} digital oscilloscope and spectrum analyser interface. 

\begin{figure*}[!th]
\centering
\includegraphics[width=0.95 \textwidth]{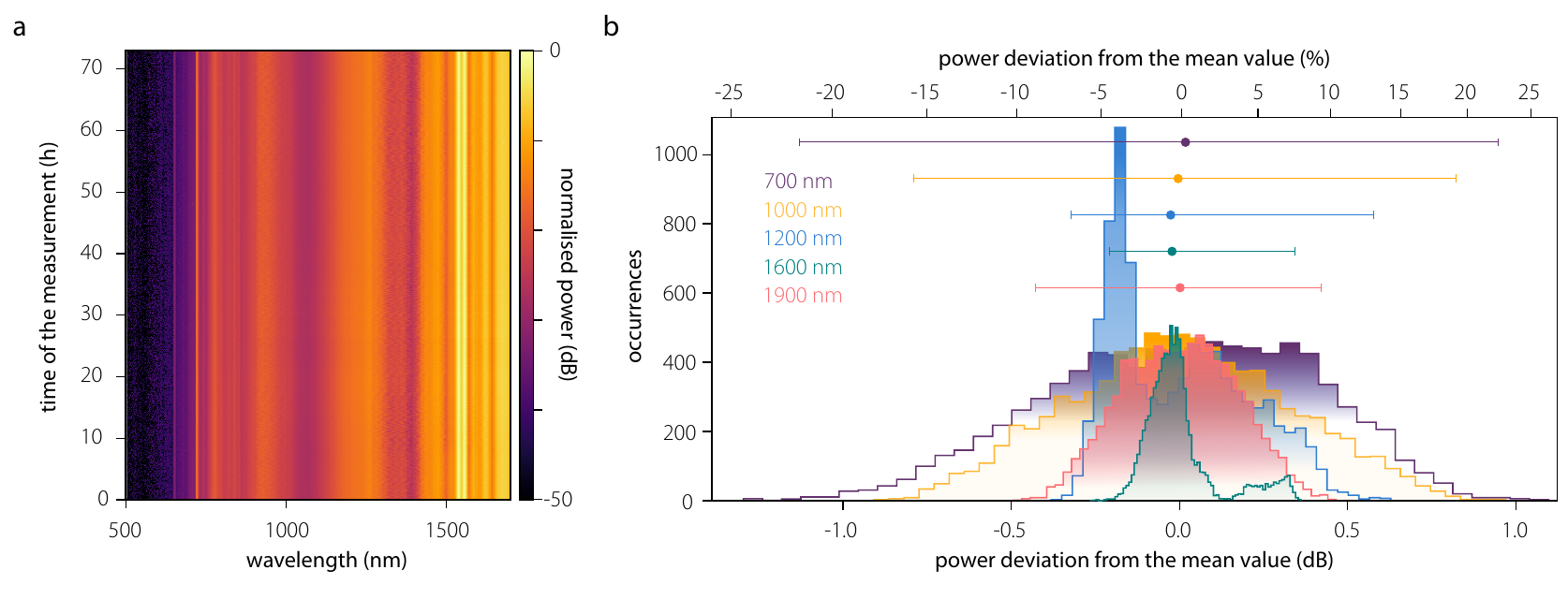}
\caption[width=1\textwidth]{\textbf{Stability of the optical supercontinuum.} \textbf{a} Long-term spectrogram showing the stability of the supercontinuum over 72 hours. \textbf{b} Histograms of the power variation from the mean value calculated for the following wavelengths: 700\,nm, 1000\,nm, 1200\,nm, 1600\,nm and 1900\,nm for the data shown in a. The horizontal lines indicate the $\pm$3\,$\upsigma$ limits for the corresponding histograms, dots mark the medians. The values are listed in Tab.\ref{tab}.\vspace{0.25cm}}
\label{fig:Spectra}
\end{figure*} 

\begin{table}[htb]
\vspace{-0.25cm}
\caption{$\pm$3\,$\upsigma$ limits, as a relative power deviation in dB, shown in Fig.\ref{fig:Spectra}b\vspace{0.2cm}}
\label{tab}
\centering
\small
\begin{tabular}{c*{4}{c}}
\hline
\hspace{0.5cm}$\uplambda$ (nm)\hspace{0.5cm} & \hspace{0.5cm}-3\,$\upsigma$\hspace{0.5cm} & \hspace{0.5cm}+3\,$\upsigma$\hspace{0.5cm} & \text{\hspace{0.5cm}median\hspace{0.5cm}} \\
 \hline\hline
 700 & -1.12 & 0.94 & 0.018 \\ 
 \hline
 1000 & -0.79 & 0.82 & -0.004 \\
 \hline
 1200 & -0.32 & 0.57 & -0.026 \\
 \hline
 1600 & -0.20 & 0.34 & -0.022 \\
 \hline
 1900 & -0.42 & 0.42 & 0.002 \\ 
 \hline
 \vspace{0.3cm}
\end{tabular}
\end{table}

A very stable hydrogen maser system supplies the 10\,MHz RF reference signal to all RF instruments in the experiment, including the FPGA board.

\section{Experimental Results}
Figure \ref{fig:setup}e showcases the f\uscript{ceo} beatnotes for both the IL (left, violet trace) and OOL (right, teal trace) systems. The beatnotes are characterised by a high signal-to-noise ratio (SNR) of 56\,dB and 45\,dB for the IL and OOL, respectively, at a resolution bandwidth (RBW) of 30\,kHz.

Once the laser is fully locked using the IL signal from the LNOI waveguide, both the IL and OOL f\uscript{ceo} were recorded without dead time or gaps by two frequency counters (Keysight 53230A) in order to verify the stability of the laser lock. Figure \ref{fig:FreqStab}a shows the f\uscript{ceo} stability measurement for over 80000 seconds (left panels) and the corresponding histograms (right panels). No phase slip was observed for both recorded data sets over this period of time. The standard deviation of the measured f\uscript{ceo} is 0.1\,mHz for the IL and 6.8\,mHz for the OOL signal. Based on these data sets the Allan deviation was calculated\cite{allantools, Harris2020} for both frequencies as shown in Fig. \ref{fig:FreqStab}b. The IL f\uscript{ceo} fractional instability of $3 \cdot 10^{-12}$ is achieved at an averaging time of 1\,s and reaches $10^{-14}$ at an averaging time of $3 \cdot 10^4$\,s. The obtainable frequency stability is likely limited by the resolution of the frequency counters as demonstrated by Dunker et al. \cite{Dunker2016}. This also implies an upper limit for the instability relative to the optical carrier frequency of $\approx 3 \cdot 10^{-19}$ and $\approx 10^{-21}$ for the averaging times of 1\,s and $3 \cdot 10^{4}$\,s, respectively. For the OOL f\uscript{ceo}, a fractional instability of $2 \cdot 10^{-10}$ is obtained at an averaging time of 1\,s and reaches $4 \cdot 10^{-13}$ at an averaging time of $3 \cdot 10^{4}$\,s, what translates to the relative instability of $3 \cdot 10^{-20}$. A similar measurement was performed for the repetition rate f\uscript{rep}, showing a standard deviation of 0.4\,mHz. However, the f\uscript{rep} was recorded with a non-zero dead-time frequency counter (HP 53132A), for which the Allan deviation cannot be calculated in a reliable manner.

In Fig. \ref{fig:FreqStab}c the phase noise is plotted at a carrier frequency of 31\,MHz for both the IL and OOL f\uscript{ceo}. The root-mean-square (RMS) integrated phase noise over the offset frequency range from 1\,Hz to 1\,MHz is 233\,mrad and 255\,mrad for the IL and OOL f\uscript{ceo}, respectively. Both measurements were also performed for the f\uscript{rep}, indicating an integrated phase noise of 1.7\,mrad.

In addition, the stability of the optical spectrum was determined over an extended period of time. In order to record the optical spectra, the light at the output of the LNOI waveguide is coupled into a 100\,$\muup$m multimode optical fibre to be recorded by an optical spectrum analyser (OSA). The data was recorded every 30 seconds for 72 hours. The evolution of the optical spectrum over time is shown in Fig. \ref{fig:Spectra}a. In order to determine the spectral stability in a quantitative manner, the power variability at selected wavelengths (700\,nm, 1000\,nm, 1200\,nm, 1600\,nm and 1900\,nm) was calculated for the entire data set and is plotted in Fig.\ref{fig:Spectra}b. The horizontal lines indicate the $\pm$3\,$\upsigma$ limits and the dots mark the median for the corresponding histograms, which are also listed in Tab.\ref{tab}. Overall, in the far blue side of the spectrum, the power deviation from the mean value does not exceed 25$\%$ and is contained within $\pm\text{15}\%$ for the majority of the spectral span. The data indicates that the setup generates highly stable optical spectra with minimal deviations over several days.
\section{Discussion}

The main limitation of the current system lies in the input coupling losses to the waveguide. The lensed fibre-to-chip interface is a convenient however not an ideal solution for coupling the light to the waveguide. As the size and shape of the modes are not perfectly matched, high losses on the input facet are observed. The measured coupling losses of approximately 6.2\,dB per facet to the waveguides with an air top cladding could be substantially improved by adding a silicon dioxide top cladding at the coupling region and designing on chip couplers acting as mode converters. Using such an approach, impressive results of only 0.5\,dB loss per facet have already been achieved with the LNOI platform\cite{Ying2021, Hu2021}. If adopted here, such a solution would further decrease the laser power required for generating an octave-spanning spectrum and for achieving self-referencing. 

In the current setup, the input fibre position is actively stabilised in order to provide a constant coupling to the waveguide by compensating for any mechanical, temperature and pressure related perturbations. From the perspective of stability and a lower complexity in particular for future applications it would be preferential to fix the input fibre using an appropriate packaging technology.

The use of a single FPGA with a flexible lock architecture for the stabilization of f\uscript{rep} and f\uscript{ceo} simplifies the setup substantially and provides a possibility to implement an orthogonal stabilisation \cite{Holman2003,Bourbeau2021} in a straight-forward way what could lead to a further decrease in the residual timing jitter.

The present setup is characterised by high stability results that show at least one order of magnitude improvement on the standard deviation of the IL f\uscript{ceo} as compared to other published works focused on direct on-chip self-referencing \cite{Hickstein2017, okawachi_chip-based_2020, Hickstein2019}. Comparing the results to dedicated systems for ultra-stable operation \cite{Fuji2005, Baumann2009, Kundermann2014}, we conclude that there is still some room for improvements.

One well-known drawback of lithium niobate is a reported relatively high sensitivity towards the photo-refractive effect\cite{kong_recent_2020, xu_mitigating_2021} and even optical damage processes\cite{Furukawa2001, Schwesyg2011}. It is uncertain to what extend photorefractive damage may limit the performance and longevity also of LNOI PICs. Within the time periods presented in this paper, and with semi-continuous illumination of different waveguides for several weeks, the optical spectra do not show any significant deviations from the initial spectrum, indicating that such detrimental effects do not pose immediate limitations for nonlinear applications at medium average input powers around 1550\,nm. However, some waveguides did show damage on the input facets that was clearly visible in electron microscopy images. While the reason for this is not yet entirely clear, the measures listed above to improve the input coupling efficiency will likely improve the durability here as well. An added top cladding would improve the thermal and mechanical stability and would protect the waveguide of contamination in this sensitive region. Additionally, larger spot-sizes at the input would decrease the local intensity and reduce the risk of related issues. For applications requiring operation with higher average powers, shorter wavelength or over longer periods of time further research is needed.

While our experiments rely on a mode-locked laser as a frequency comb source, the results could also pave the way for efficient self-referencing of Kerr frequency combs in the future. In this case, the broadband spectrum can be generated directly on the PIC from a continuous wave laser \cite{brasch_photonic_2016, pfeiffer_octave-spanning_2017, briles_hybrid_2021, Gong2020}. However, in contrast to earlier demonstrations \cite{jost_counting_2015, brasch_self-referenced_2017, spencer_optical-frequency_2018, Liu2021}, the required frequency doubling could be realized on the same PIC. Using an additional external lasers, one step in this direction was taken very recently using an aluminum nitride PIC \cite{Liu2021}.

\section{Conclusions}
We have demonstrated a compact and stable RF-to-optical link employing lithium niobate on insulator waveguides. Owing to the high $\chiup^{(2)}$ and $\chiup^{(3)}$ nonlinearities of the lithium niobate, injecting low energy pulses ($\sim$ 140\,pJ) into the waveguides results in a broadband supercontinuum overlapping with the second harmonic, which allows for direct on-chip f-2f interferometry. We measure the in-loop f\uscript{ceo} beatnote with a high SNR of 56\,dB, which is the basis for a high quality full stabilisation of the frequency comb using a single FPGA operating on open source software. The laser lock showcases an excellent stability, demonstrating an upper limit on the Allan deviation of $3 \cdot 10^{-12}$ at an averaging time of 1\,s and an integrated phase noise of 234\,mrad. The out-of-loop monitoring based on a standard f-2f interferometer built from fiber optics and bulk optics complements the metrological characterisation of the system. These results show that lithium niobate on insulator is a suitable platform for applications in metrology and nonlinear optics. Compact and efficient LNOI-based self-referencing setups could replace bulky standard f-2f interferometers for laser stabilisation and could be used in a variety of applications requiring a broadband optical spectrum and a very good phase noise performance.

\section*{Funding Information}
This work was supported by the Swiss National Science Foundation and Innosuisse via the Bridge Discovery Grant 40B2-1\_176563, the Swiss National Science Foundation with Grant CRSII5\_193689 as well as the Canton of Neuchâtel.

\section*{Acknowledgments}
We would like to acknowledge the work of the staff of the CSEM cleanroom facilities, in particular Yves Petremand, Mehretab Amine and Patrick Surbled. We would also like to thank the many authors and contributors of the open source software community for their great work on tools such as Python, NumPy\cite{Harris2020}, Matplotlib\cite{Hunter2007} and Inkscape.

\section*{Author Declarations}
CSEM offers LNOI-related design, fabrication, testing and integration services. The authors have no other conflicts to disclose. 

\section*{Data availability}
The data that supports the findings of this study are openly available in Zenodo at 10.5281/zenodo.5571451.

\selectlanguage{english}
\bibliography{LNLongTerm.bib}

\end{document}